\documentclass{elsart3p}
\usepackage{graphicx,amssymb}
\journal{Nuclear Instruments and Methods in Physics Research B}
\begin{document}
\begin{frontmatter}

\title{Scintillation light produced by low-energy beams\\
of highly-charged ions}

\author[a]{M. Vogel\corauthref{cor}},
\corauth[cor]{Corresponding author.} \ead{m.vogel@gsi.de}
\author[a]{D.F.A. Winters},
\author[b]{H. Ernst},
\author[c]{H. Zimmermann}, and
\author[a]{O. Kester}
\address[a]{GSI, Planckstrasse 1, D-64291 Darmstadt, Germany}
\address[b]{Johannes-Gutenberg-Universit\"at Mainz, Staudingerweg 7, D-55099 Mainz, Germany}
\address[c]{Ludwig-Maximilians-Universit\"at, Schellingstrasse 4, D-80799 M\"unchen, Germany}

\begin{abstract}
Measurements have been performed of scintillation light
intensities emitted from various inorganic scintillators
irradiated with low-energy beams of highly-charged ions from an
electron beam ion source (EBIS) and an electron cyclotron
resonance ion source (ECRIS). Beams of xenon ions Xe$^{q+}$ with
various charge states between $q$=2 and $q$=18 have been used at
energies between 5\,keV and 17.5\,keV per charge generated by the
ECRIS. The intensity of the beam was typically varied between 1
and 100 nA. Beams of highly charged residual gas ions have been
produced by the EBIS at 4.5 keV per charge and with low
intensities down to 100 pA. The scintillator materials used are
flat screens of P46 YAG and P43 phosphor. In all cases,
scintillation light emitted from the screen surface was detected
by a CCD camera. The scintillation light intensity has been found
to depend linearly on the kinetic ion energy per time deposited
into the scintillator, while up to $q$=18 no significant
contribution from the ions' potential energy was found. We discuss
the results on the background of a possible use as beam
diagnostics e.g. for the new HITRAP facility at GSI, Germany.
\end{abstract}

\end{frontmatter}

\section{Introduction}
Beams of heavy highly-charged ions are typically produced at
energies in the high keV/u to MeV/u range. Correspondingly, the
energy deposition of such a beam into a solid target, e.g. a
scintillation screen, is high and the subsequent light emission
intense. At the HITRAP facility \cite{hitrap,hitrap2}, beams of
highly-charged ions will be produced by stripping, using the GSI
accelerator facility. Stripping of heavy ions to high charge
states requires beam energies of several hundred MeV per nucleon.
At HITRAP these ions will subsequently be decelerated to energies
of only a few keV per charge. Beam intensities at HITRAP will be
below a $\mu$A during a bunch of $\mu$s duration. It will be
necessary to have beam diagnostic intrumentation at hand which is
able to detect such low-energy particle beams. Especially, the low
kinetic energy deposition into the scintillator and its possible
charging up, leading to a subsequent deflection of the low-energy
beam, may be obstacles for optical detection of charged particle
beams.

Numerous scintillation experiments have been performed with a wide
variety of scintillator materials, however mostly photon or
electron beams have been used to stimulate scintillation
\cite{leo}. Ion beam experiments have been predominantly performed
with intense beams of MeV/u kinetic energies, see for example
\cite{p1,p2,p3}. For low-energy beams both experimental \cite{a1}
and theoretical \cite{a2} results are available, however no high
charge states have been considered. Information on scintillation
light produced by low-intensity, highly-charged ion beams in the
energy region of few keV per charge is sparse and therefore
requires a more detailed study. Generally, it is known that for
ions the scintillation light yield is roughly 50 to 70\% of that
for electrons and about 25 to 50\% of that for photons \cite{leo}.
Scintillator materials like the presently used P43 phosphor screen
emit on average 36 photons per keV deposited kinetic energy, P46
(YAG) emits about half of that \cite{leo}.

In the following, we present the results of an experimental study
of scintillation light produced by low-energy highly-charged ion
beams impinging on P43 and P46 scintillator screens.

\begin{figure}[!bt]
\begin{center}
\centering
\includegraphics[width=7.5cm]{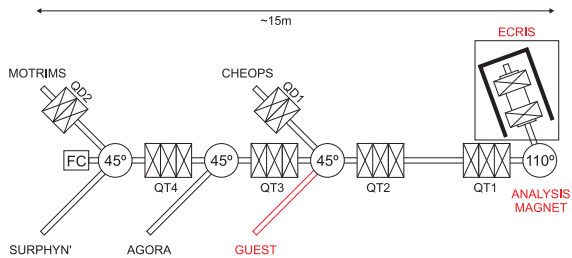}
\caption{ECR ion source setup at KVI, Groningen with its "guest"
port, where the present experiments have been performed.}
\label{fig1}
\end{center}
\end{figure}

\section{Experimental setup and procedure}
The measurements have been performed using an Electron Cyclotron
Resonance Ion Source (ECRIS) \cite{ecr1,ecr2} setup at the KVI,
Groningen and the MAXEBIS setup at GSI, which is described in
detail in \cite{icis05}. Details of the KVI-ECRIS have been given
in \cite{ecr3}. In case of the ECRIS, highly-charged ions are
continuously produced and transferred to the anaysis magnet which
is a mass-over-charge ($m/q$) filter. The ion species of interest
is selected by an appropriate choice of the field strength in the
analysis magnet. The beam energy is defined by the potential
applied to the source. Values up to 20\,kV can be chosen with an
uncertainty of about 50\,V. Ion optical elements then guide the
beam to the "guest" port shown in figure \ref{fig1}.

The MAXEBIS delivers a pulsed beam and therefore the CCD-camera
was triggered by the signal of the extraction pulser. Depending on
the confinement time of the ions in the EBIS, the repetition rate
is typically of the order of 10 to 100 Hz. The ion pulse length
corresponds to 30 - 50 $\mu$s, typical beam energies are of the
order of a few keV per charge.

\begin{figure}[!tb]
\begin{center}
\centering
\includegraphics[width=7.5cm]{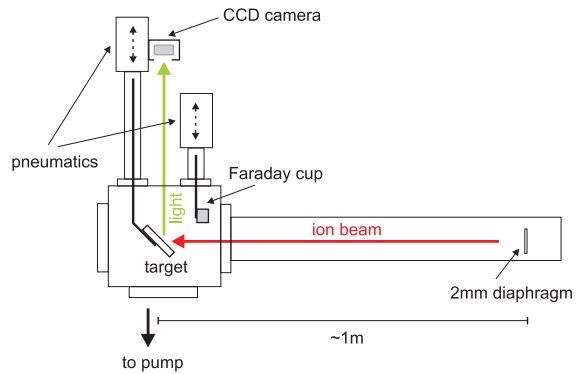}
\caption{Scheme of the experimental setup. For details see text.}
\label{fig2}
\end{center}
\end{figure}

The scintillator screens are irradiated by the ion beam under an
angle of 45$^o$ and a CCD camera (model Basler A311f) monitors the
scintillation light emitted from the screen perpendicular to the
beam axis, see figure \ref{fig2}. A Faraday cup (30\,mm diameter)
with secondary electron suppression can be moved into the beamline
for a measurement of the beam current. The uncertainty of this
measurement is estimated to be about 1\,\%. However, temporal
fluctuations of the beam intensity can amount up to 10\,\% of the
measured value.

The P43 target is an aluminium disk of 70\,mm diameter coated with
about 40-50\,$\mu$m of amorphous Gd$_2$O$_2$S:Tb. It emits light
between 360\,nm and 680\,nm with a spectral peak wavelength of
545\,nm (green) and has an average decay time (90\% to 10\%
intensity) of 1.0\,ms. The P46 YAG is a Y$_3$Al$_5$O$_{12}$:Ce
crystal which also emits green light with a spectral peak
wavelength of 550\,nm and has an average decay time of 70\,ns.
Additionally, we have used a P46 YAG screen identical to the first
one, but coated with 1\,$\mu$g/cm$^2$ of aluminium to build a
conducting surface and possibly prevent it from charging up by the
ion beam.

In all cases it may be assumed that the ions deposit their
complete kinetic energy into the target, i.e. that the
scintillation signal represents the beam characteristics well.
This is corroborated by application of the Bethe-Bloch equation
for the energy loss of the ion in the target material \cite{leo},
presently leading to loss coefficients of the order of GeV/mm and
correspondingly to extremely low penetration depths which are well
below the scintillator thickness. Also, the present scintillators
are highly transparent at their emission wavelengths such that
nearly all produced light is emitted. At the low beam intensities
presently used, saturation effects of the scintillators are not
expected. The experiments have been performed at room temperature
(T$\approx$300\,K) and at a residual gas pressure of a few
$10^{-7}$\,mbar. Due to the low ion beam currents and energies
used, heating of the target is not expected. The temporal
stability of the scintillation light signal can be estimated from
the measured fluctuations to be well within 10\% of the observed
intensity within a time interval of about 15 minutes.

\begin{figure}[!tb]
\begin{center}
\centering
\includegraphics[width=3.5cm]{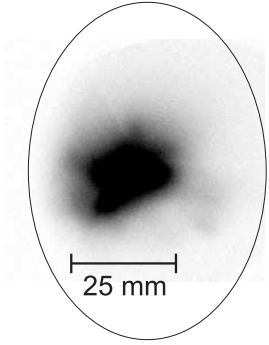}
\caption{Beam spot on the P46 scintillator screen for 30 nA of
Xe$^{7+}$ at 10\,keV/q. For the sake of demonstration, the picture
has been color inverted and shows the screen border.} \label{fig3}
\end{center}
\end{figure}

\section{Results}
Figure \ref{fig3} shows a typical picture of the beam spot on the
scintillator as seen by the CCD camera. For the sake of
demonstration, this picture has been taken at the highest camera
sensitivity and is colour inverted. The ellipse represents the
scintillator surface as seen by the camera under the 45$^o$ angle
(see figure \ref{fig2}). In the further evaluation, the amount of
scintillation light ("intensity") is determined from unsaturated
greyscale pictures (646 x 494 pixels) as the sum of all greyscale
values. The amount of background light has been measured and
subtracted from all images before further analysis.

\begin{figure}[!tb]
\begin{center}
\centering
\includegraphics[width=7.5cm]{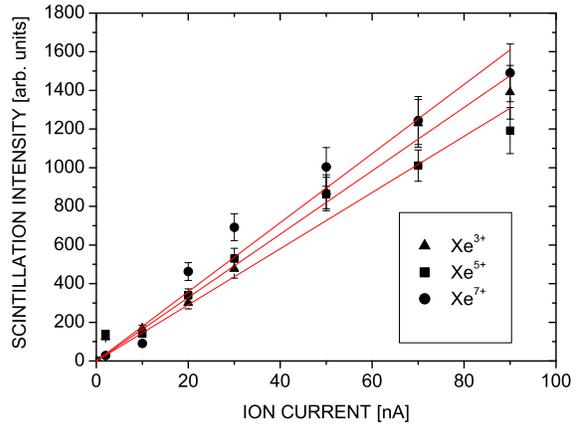}
\caption{Detected scintillation light from the P46 target as a
function of the ion beam current for Xe$^{3+,5+,7+}$ at an energy
of 10\,keV/q. The line is a weighted linear fit to the data.}
\label{fig4}
\end{center}
\end{figure}

The scintillation light intensity is found to depend linearly on
the ion beam current (see figure \ref{fig4}). This is in agreement
with both theoretical considerations and previous experimental
results on singly and multiply charged ions at high energies
\cite{p1,p2,p3,anto,blas,knol}. For fixed beam parameters the
kinetic energy deposited into the scintillator is proportional to
the beam current. As an example, figure \ref{fig4} shows the
intensity from the P46 target as a function of the ion beam
current for Xe$^{3+}$, Xe$^{5+}$ and Xe$^{7+}$ ions at an energy
of 10\,keV/q. The dependence is linear as suggested by the fits.
This linear relationship holds true for all ion species under
observation and for all kinetic energies used in the present
investigation.

\begin{figure}[!bt]
\begin{center}
\centering
\includegraphics[width=7.5cm]{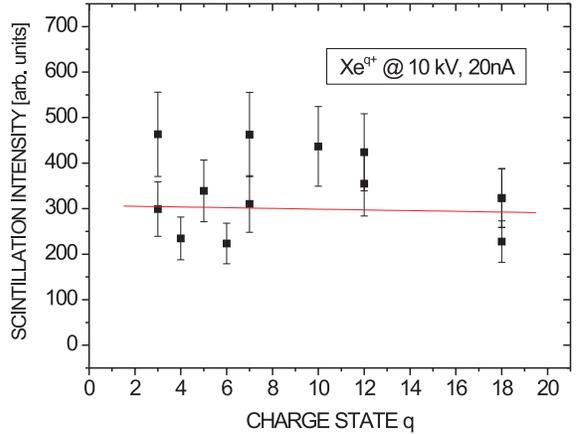}
\caption{Scintillation light intensity from the P46 screen as a
function of the ion charge state $q$ measured with 20 nA beams of
Xe$^{q+}$ ions at 10\,keV/q. The line is a weighted linear fit to
the data.} \label{fig5}
\end{center}
\end{figure}

From figure \ref{fig4} it can also be concluded that there is no
significant effect of the charge state on the intensity. To
quantify this further, the charge-state dependence has been
investigated in more detail. Figure \ref{fig5} shows the relation
between charge state and intensity for 20\,nA beams of Xe$^{q+}$
ions at an energy of 10\,keV per charge on the P46 target.

The scatter of this data exceeds the 10\,\% statistical
uncertainty given by beam intensity fluctuations (see above). This
is attributed to surface-related effects, such as surface
contamination (residual gas) and ion beam sputtering of the
surface (cleaning). However, within the scatter of the data, there
seems to be no obvious dependence on the charge state. Since the
ion beam current $I$ is kept constant, an increase in charge state
$q$ leads to a corresponding decrease of the ion number current
proportional to $1/q$. At the same time, since the acceleration
voltage is fixed, the kinetic energy per particle increases
proportional to $q$, such that the total amount of deposited
energy per time is unchanged.

An important number for practical use of scintillators as beam
diagnostics is the "threshold current". This is the minimum
current required to produce an optically detectable
scintillation signal.

\begin{figure}[tb]
\begin{center}
\centering
\includegraphics[width=7.5cm]{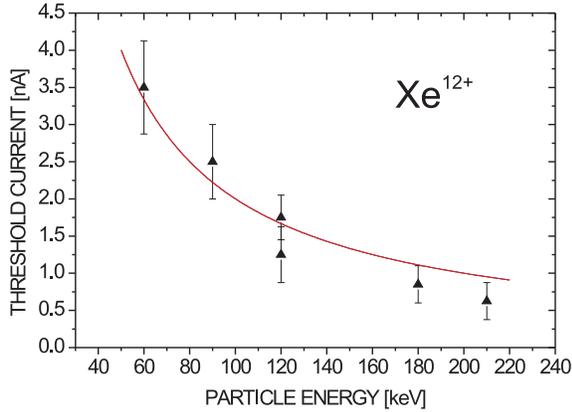}
\caption{Ion beam current threshold as a function of particle
energy for a given particle, here Xe$^{12+}$. The line is a $1/E$
fit to the data.} \label{fig6}
\end{center}
\end{figure}

Figure \ref{fig6} shows the dependence of the threshold current on
the particle energy for Xe$^{12+}$. The threshold current shows a
$1/E$ dependence, where $E$ is the kinetic energy of the particle.
This can be understood as follows: With more kinetic energy per
particle, less particles are needed to produce the same intensity.
For fixed $q$, this means that the threshold current decreases.

All ECRIS threshold data was taken at the highest CCD camera
sensitivity. This implies that $3.65 \times 10^7$ photons per
observation time of 80\,ms are required for a non-vanishing
signal.

With the MAXEBIS, the dc-threshold current has been measured to be
100 pA for a focused C$^{4+}$-beam at a kinetic energy of 4.5
keV/q. However, due to the focused beam (spotsizes of only a few
mm), the apparent threshold current is lower than for the ECRIS
measurements by the ratio of the spot areas. This is also a
practical means to lower the threshold current, as long as the
spotsize can be controlled well and the focusing does not lead to
saturation of the scintillation material.

The information on the threshold current can, together with the
known camera sensitivity and beam parameters, be used to extract
information on the photon yield $n$ of the scintillator. At the
threshold current $I$, the particle current is $I_n=I/qe$, where
$e$ is the elementary charge and $q$ is the charge state of the
ion. The kinetic energy deposited per second into the scintillator
is $E_{kin}=q I_n U$, where $U$ is the ion acceleration voltage.
We assume that at this value, the light emission is just at the
camera threshold, i.e. $3.65 \times 10^7$ photons are detected
within 80\,ms.

We define the photon yield of the scintillator as
\begin{equation}\nonumber
n=\frac{\textrm{number of emitted photons}}{\textrm{kinetic
energy deposition}}.
\end{equation}
Expressing this in terms of the experimental variables yields
\begin{equation}\nonumber\label{eqq}
n=\frac{N}{A B E_{kin}}=\frac{N e}{A B I U}.
\end{equation}
Here $A=5 \times 10^{-4}$ is a solid angle correction for the
fraction of emitted photons seen by the camera, given by
$A=r^2/(4d^2)$ where $r=0.02$\,m is the camera lens aperture
radius and $d=0.45$\,m is the screen-lens distance. $N=3.65 \times
10^7$ is the threshold photon number and $B=0.08$ is the fraction
of a second during which photons are detected.

\begin{figure}[tb]
\begin{center}
\centering
\includegraphics[width=7.5cm]{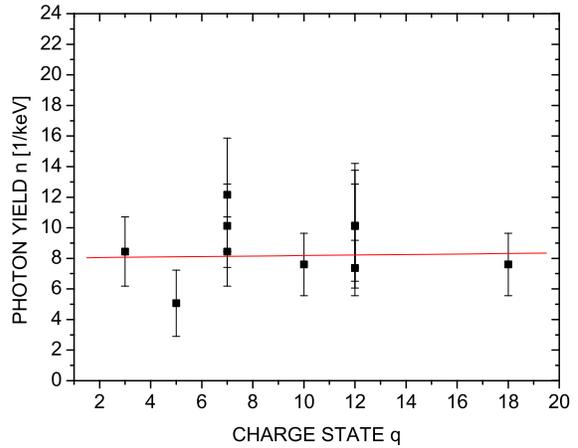}
\caption{Photon yield $n$ in units of 1/keV as a function of the
charge state $q$ of the xenon ions. The line is a linear fit to
the data.} \label{fig7}
\end{center}
\end{figure}

Figure \ref{fig7} shows the photon yield $n$ in units of 1/keV as
a function of the charge state $q$. The line is a linear fit to
the data and its slope is in agreement with zero. Thus, the photon
yield is found not to depend on the ions' charge state, but to
have a constant mean value of $n=8.1 \pm 1.8$.

This is in agreement with the independence of equation (\ref{eqq})
on $q$ and with the above statement that for ions the photon yield
of P46 is between 25\% and 50\% of 18 photons per keV of deposited
energy. Furthermore, it corroborates the above discussed
independence of the scintillation signal on the ions' charge state
(see figure \ref{fig5}).

Under certain conditions, the P43 target showed charging phenomena
expressed by a continuous increase in emitted scintillation light
from zero to a maximum value and an intense subsequent flash. This
has been observed exclusively for multiply-charged particles of
energies of and above 10\,keV per charge. The light intensity
during the flash is about two orders of magnitude higher than the
brightest regular intensity before the flash. The duration of the
flash is well below a single frame of the CCD camera, which is
about 80\,ms. As long as the beam hits the target, this phenomenon
is repeated at a constant flashing rate depending on the ion beam
parameters.

\begin{figure}[tb]
\begin{center}
\centering
\includegraphics[width=7.5cm]{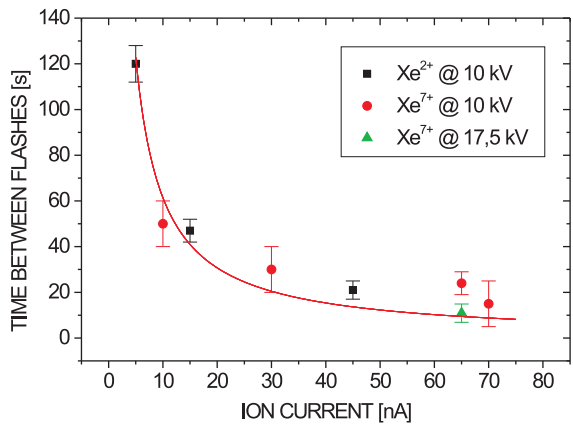}
\caption{Average time between observed flashes at the P43 target
as a function of ion beam current $I$. The line is a 1/$I$ fit to
the data (see text).} \label{fig8}
\end{center}
\end{figure}

Figure \ref{fig8} shows the average time between observed flashes
of the P43 target as a function of the ion beam current for
different multiply-charged ions at different ion energies. The
necessary charging time between flashes shows a $1/I$ dependence,
as can be seen from the fit in figure \ref{fig8}. In contrast, the
P46 YAG screens have not shown any of these phenomena under the
same conditions. The flashing phenomenon might, however, be used
for a measurement of the ion beam current in situations where
light detection is difficult and it is only possible to observe
the very bright flashes. In this kind of situation, the flashing
rate could be used as a measure of the ion beam current.

Apparently, the thin amorphous layer of P43 on eloxy aluminium has
a substantially higher electrical resistance than the P46 crystal,
for which no charging phenomenon was observed for beam currents up
to one $\mu$A. In contrast to the observations made in
\cite{sieber}, no temporal decay of the scintillation efficiency
could be observed for both P43 and P46, which appear to keep
constant efficiencies over beam times of hours. However, P43
showed similar surface anomalies as those described in
\cite{sieber}, i.e. a local change in colour from white to brown,
but without any macroscopically observable change in structure or
in scintillation efficiency. This change in colour may be
attributed to the flashing phenomenon, which presumably is
connected with high currents which may give rise to changes in the
surface composition. In situations with high-intensity high-energy
beams, it was found that carbon from cracked residual gas
molecules had been adsorbed to the target surface thus changing
its composition and colour \cite{tob}. Although the presently used
beam intensities and energies are several orders of magnitude away
from those conditions, it may still be assumed that the very
intense flashes across the screen surface are connected with
temperatures high enough to crack residual gas molecules which
then are adsorbed onto the surface.

The metal-coated P46 screen was found to show no observable
scintillation light for any of the beam parameters used. Thus, a
number of high-current measurements in the $\mu$A region were
performed using beams of He$^+$ and O$^{2+}$. However, up to beam
currents of 1 $\mu$A, no scintillation was detected. The metal
coating is believed to be thin enough (about 10 monolayers) for
the ions to penetrate even at low energies. Yet, scintillation was
hindered by the metal coating for unknown reasons. This behaviour
is unexpected and requires a more detailed study.

\section{Summary and conclusion}
We have performed scintillation light measurements with continuous
and pulsed beams of low-energy, highly-charged ions on P43 and P46
scintillator screens. While the metal coated P46 screen did not
show any observable scintillation signal for any beam current up
to 1 $\mu$A, both uncoated P43 and P46 screens showed signals for
beam currents above roughly 1\,nA for continuous beam from the
ECRIS and about 100 pA for focused and pulsed beams from the
MAXEBIS. The intensity of this signal has been found to increase
linearly with the beam current, in agreement with previous
findings at high energies and theoretical considerations. This
behaviour is found to be independent of the charge state of the
ions, at least up to Xe$^{18+}$. The amount of produced
scintillation light thus seems to depend only on the amount of
kinetic energy deposited into the scintillator and not on the
charge state of the ions. This holds true both for the observed
intensity and the photon yield per deposited energy. The P43
screen shows wear and charging phenomena which may obstruct its
use under certain conditions, i.e. for high particle energies or
intense beams. The uncoated P46 seems to be best suited for
diagnostics of low-energy beams of highly-charged ions, e.g. at
HITRAP. This is especially due to the fact that P46 produces
sufficient intensity while not showing any signs of charging up
even when highly charged ions are used. It is yet unclear if the
present results apply also to ions with much higher charge states,
since secondary effects due to the high potential energy cannot be
ruled out. Also, potential sputtering \cite{hay,dude} could become
significant for very highly charged ions. This needs experimental
clarification when such ion beams become available. The fast
scintillation decay of P46 may evoke the need for a more sensitive
camera system than presently used when a pulsed beam is
considered. This requirement, however, can easily be met.

\section{Acknowledgements}
We are grateful for the fruitful collaboration with T.
Schlath\"olter and R. Hoekstra from KVI, Groningen, Netherlands.
We also thank T. Hoffmann and P. Forck (GSI) for helpful
discussions.

\end{document}